\documentclass{report}                         
\usepackage[backref,colorlinks=true]{hyperref} 
\input{epsf}                                   
\begin{document}                               

\bigskip
\bigskip

\chapter*{\bf Non-inertial quantum mechanical fluctuations}
\vskip 0.5cm
{ {\bf Haret Rosu }
\\[15pt]
Guanajuato University\\
Leon\\
Guanajuato\\
Mexico\\
E-mail:  rosu@ifug3.ugto.mx }
\vskip 2 cm
\noindent {\bf Abstract:} 
{\it Zero point quantum fluctuations as seen from non-inertial
reference frames are of interest for several reasons. In particular,
because phenomena such as Unruh radiation (acceleration radiation) and
Hawking radiation (quantum leakage from a black hole) depend
intrinsically on both quantum zero-point fluctuations and some
appropriate notion of an accelerating vacuum state, any experimental
test of zero-point fluctuations in non-inertial frames is implicitly a
test of the foundations of quantum field theory, and the Unruh and
Hawking effects.}

\vskip 1cm

\noindent
{\sl To appear as a chapter in the book {\em Artificial Black Holes} (World
Scientific) edited
by M. Novello, M. Visser, and G. Volovik.}

\vskip 2 cm
\def\ie{{\sl i.e.}}
\def\eg{{\sl e.g.}}
\def\d{{\mathrm{d}}}
\section{Introduction}
\setcounter{equation} {0}

Analog or not, the ultimate goal of the physics described in this book
is to find clear evidence for gravitational and non-inertial vacuum
radiation.  Recognized as some of the most important paradigms of
present-day theoretical physics, the Hawking and Unruh effects are (as
yet) not much more than academic results which, because of the scales
of the required energies/ accelerations/ masses, are not easy to
implement in real laboratory experiments.  Indeed, from the
experimental standpoint, they might seem to be merely exotic
interpretations for what could be explained by far more mundane
physical effects in quantum electrodynamics, quantum optics, and
hydrodynamics.

As counterpoint, in the case of the Unruh effect the so-called
detector method provides a well-defined radiation pattern that may be
thought of as vacuum noise, and should be kept under consideration for
possible experimental detection in clean analog experiments.  In this
chapter, old results of Letaw on scalar vacuum radiation patterns are
used to emphasize the radiometric nature of the various Frenet--Serret
invariants for certain classes of ``stationary'' worldlines. This
formalism is an extension of, and alternative to, the usual notion:
that of adopting the thermal interpretation of the vacuum excitations
as seen by a uniformly accelerated quantum detector (Unruh's
interpretation).

I focus next on the electromagnetic vacuum noise, surveying the
Hacyan-Sarmiento approach for calculating physical quantities in the
electromagnetic vacuum.  The application of this approach to circular
worldlines led Mane to propose the identification of Hacyan--Sarmiento
zero-point radiation with the ordinary synchrotron radiation; but here
I provide some simple counter-arguments.  I also briefly discuss Bell
and Leinaas' proposal of considering electrons in storage rings as
prototypes for an Unruh--DeWitt spin polarization detector.  In the
final section, I sketch the similarity between the Unruh effect and
the so-called ``anomalous Doppler effect''. A separate observation of
the latter would mean a confirmation of the possibility of the first.

\section{Vacuum Field Noise --- VFN}

The quantum vacuum field noise (VFN)~\cite{T} that is recorded by a
detector moving along some classical trajectory will in general depend
on that trajectory.  For certain restricted classes of worldline
trajectories $x^{\mu}(s)$, most usefully parametrized in terms of
proper time $s$, the observed power spectrum is stationary
(time-independent).  Experimental observation of the power spectrum of
the vacuum field noise is then an important diagnostic tool that can
be inverted to extract information about the form of the trajectory
--- specifically, the curvature invariants (Frenet--Serret invariants)
of the worldline.  As usual, I model an idealized detector as a simple
two-level quantum system [usually known as an Unruh--DeWitt detector].
For scalar quantum field vacua there are six broad classes of
trajectory that lead to stationary noise spectra. Basic results were
derived by Letaw~\cite{L} some time ago, and are reviewed below.  They
might be of direct experimental interest in the acoustic analogy.  On
the other hand, one should also keep in mind that non-stationary
vacuum noises are not completely beyond experimental reach, and can be
analyzed by related mathematical methods which I briefly comment on.

\subsection{The detector method in quantum field theory} 
\label{S:udw}

For the idealized Unruh--DeWitt detector, the interaction between the
detector [endowed with a monopole moment $Q(s)$\,] and the scalar
field $\phi(s)$ is described by
\begin{equation}
L_{int}=\lambda \; Q(s)\; \phi(s),
\end{equation}
where $\phi(s)=\phi(x(s))$ is the field along the worldline of the
detector and $\lambda$ is a small coupling constant that I re-scale to
$\lambda=1$ (since it is not important for current considerations).
Detecting particles in the Unruh--DeWitt apparatus requires one to
adopt adiabatic switching appropriate for the perturbative approach.
At $s=-\infty$, the detector is in the ground-state $|E_0\rangle$ and
the field is in the Minkowski vacuum $|0_{M}\rangle$. After the
detector-field interaction is switched on, the detector would not
remain in the state $|E_{0}\rangle$, but would make a transition to
$|E_{1}\rangle$. It is said that the detector ``detects'' some
particles. Then, the transition amplitude for the detector field
system to be found in $|E_1,\psi\rangle$ at $s =+\infty$ is given by
first-order perturbation theory as
\begin{equation}
A=i\left\langle E_1,\psi\left|
\int _{-\infty}^{\infty} \d s \; Q(s)\;\phi (s)
\right|E_{0},0_{M}\right\rangle.
\end{equation}
In order for first-order perturbation theory to apply one has to
assume that the matrix element of $Q$ is sufficiently small.  On the
other hand, from the time evolution of the operator $Q$ in the
Heisenberg picture
\begin{equation}
Q(s)=e^{iH_{D}\;s /\hbar}\; Q(0)\; e^{-iH_{D}\;s /\hbar},
\end{equation}
where $H_{D}$ is the detector Hamiltonian, one immediately obtains
\begin{equation}
A=i \langle E_1|Q(0)|E_{0}\rangle 
\int _{-\infty}^{\infty} \d s \; 
e^{i(E_1-E_0)s /\hbar}\; \langle \psi|\phi (s)|0_{M}\rangle .
\end{equation}
After summation over all final states of the field $|\psi\rangle$, the
transition rate, \ie, the transition probability per unit proper time
from $E_0$ to $E_1$ is
\begin{equation}
\frac{\d P}{\d s}= |\langle E_1|Q(0)|E_{0}\rangle|^2 \; S(\omega) ,
\end{equation}
where $\omega=(E_1-E_0)/{\hbar}$ and
\begin{equation}
S(\omega)=\int _{-\infty}^{\infty} \d(s - s')
\; e^{-i\omega (s - s')} \; g(s - s').
\end{equation}
The integrand $g$ is the Minkowski vacuum expectation value of the
autocorrelation function (the Wightman function)
\begin{equation}
g(s-s')=\langle 0_{M}|\phi(s)\, \phi(s')|0_{M}\rangle.
\end{equation}
Thus, $S$ looks like a response function (or power spectrum) and $g$
as the ``quantum noise" in the Minkowski vacuum along the worldline
$x(s)$.  The peculiar feature of this argument is that the quantum
detector performs an ``up'' transition and at the same time sees
(`emits') a `radiation' spectrum.  From the phenomenological point of
view such a situation can also be encountered in the case of the
anomalous Doppler effect (ADE) as has been remarked by Frolov and
Ginzburg~\cite{FG} (see Section \ref{S:ade} below).

\subsection{Six types of stationary scalar VFN}  

In general, the scalar quantum field vacuum does not possess a
stationary vacuum excitation spectrum (abbreviated as SVES) for all
types of classical relativistic trajectories on which the
Unruh--DeWitt detector could move.  Nevertheless, linear uniform
acceleration is {\em not} the only case with that property.  This was
shown by Letaw, who extended Unruh's considerations, obtaining six
broad classes of worldlines with SVES for an Unruh--DeWitt monopole
detector (SVES-1 to SVES-6, see below). The line of argument is the
following: The Unruh--DeWitt detector is effectively immersed in a
scalar bath of vacuum fluctuations.  Its rate of excitation is
determined by the energy spectrum of the scalar bath, which can be
expressed as the density of states times a cosine-Fourier transform of
the Wightman correlation function of the scalar field. Since the
Wightman function is directly expressed in terms of the inverse of the
geodesic interval what one needs to calculate is a Fourier transform
of the inverse of the geodesic interval $\int \d s =
\int\sqrt{\d x_{\mu}^2}$.  Moreover, stationarity means that the
Wightman function depends only on the proper time interval.

As shown by Letaw, the stationary worldlines are solutions of some
generalized Frenet--Serret (FS) equations on which the condition of
constant curvature invariants is imposed. That is, one is interested
in worldlines of constant curvature $\kappa$, torsion $\tau$, and
hyper-torsion (bi-torsion) $\nu$, respectively. These curvature
invariants can be easily built from the tangent, normal, and binormal
vectors and their derivatives. They have physical interpretation in
terms of the observer's acceleration and angular velocities. Notice
that one can employ other frameworks, such as the Newman-Penrose
spinor formalism as recently invoked by Unruh~\cite{un}, but the
Frenet--Serret framework is in overwhelming use throughout physics. It
is worth remarking that before Letaw, the Frenet--Serret invariants
have been discussed by Honig {\em et al}~\cite{Ho} in their study of
the motion of charged particles in homogeneous electromagnetic fields.
Honig {\em et al} discovered an interesting connection with the two
Lorentz invariants of the electromagnetic field, $E^2-H^2$ and
$\vec{E}\cdot \vec{H}$:
\begin{equation} \label{Lo1}
E^2 - H^2 \propto\kappa ^2-\tau ^2 -\nu ^2,
\end{equation}
\begin{equation}
\vec{E}\cdot \vec{H} \propto\kappa \; \nu.
\end{equation}
It is amusing to note that chirality (handedness) in the
electromagnetic sense ($\vec{E}\cdot \vec{H} \neq 0$) is proportional
to chirality in the worldline sense (nonzero hyper-torsion; $\nu\neq
0$.)

The six stationary scalar VFN can be classified according to the
curvature scalars of the corresponding worldlines:
\begin{enumerate}
\item
\underline{$\kappa =\tau=\nu=0$} $\rightarrow$
{\em inertial, uncurved worldlines (constant velocity)}. 

\bigskip

SVES-1 is a trivial cubic spectrum
\begin{equation}
S_1(E)=\frac{E^3}{4\pi ^2}.
\end{equation}
This can be interpreted as a vacuum state of zero point energy $E/2$
per mode, with density of states $E^2/2\pi ^2$.

\item
\underline{$\kappa \neq 0$, $\tau=\nu=0$}  $\rightarrow$
{\em hyperbolic worldlines (constant rectilinear acceleration)}. 

\bigskip

SVES-2 is Planckian allowing the interpretation of $\kappa/2\pi$ as
`thermodynamic' temperature. In the dimensionless variable $\epsilon
_{\kappa}=E/\kappa$ the vacuum spectrum reads
\begin{equation}
S_2(\epsilon _{\kappa})
=\frac{\epsilon _{\kappa}^{3}}{2\pi ^2(e^{2\pi\epsilon _{\kappa}}-1)}.
\end{equation}
The physically observed spectrum would be a linear combination of
SVES-1 and SVES-2.

\item
\underline{$|\kappa|<|\tau|$, $\nu=0$, $\rho ^2=\tau ^2-\kappa ^2$}
$\rightarrow$ {\em helical worldlines}. 

\bigskip

SVES-3 is a complicated analytic function corresponding to the case 4
below only in the limit $\kappa\gg \rho$
\begin{equation}
S_3(\epsilon _{\rho})\stackrel{\kappa/\rho\rightarrow \infty}
{\longrightarrow} S_4(\epsilon _{\kappa}).
\end{equation}
Letaw plotted the numerical integral $S_3(\epsilon _{\rho})$, where
$\epsilon _{\rho}=E/\rho$, for various values of $\kappa/\rho$.

\item
\underline{$\kappa=\tau$, $\nu=0$}
$\rightarrow$ {\em the spatially projected worldlines are the
``semicubical parabolas'', $y\propto\kappa \; x^{3/2}$, containing a
cusp (at $x=0$) where the direction of motion is reversed}.

\bigskip

SVES-4 is analytic, and since there are two equal curvature invariants
($\kappa = \tau$) one can use the dimensionless energy variable
$\epsilon _{\kappa} = E/\kappa$
\begin{equation}
S_{4}(\epsilon _{\kappa})= \frac{\epsilon _{\kappa}^{2}}{8\pi ^2 \sqrt{3}}
e^{-2\sqrt{3}\epsilon _{\kappa}}.
\end{equation}
It is worth noting that $S_4$, being a monomial times an exponential,
is rather close to the Wien spectrum 
$S_{W}\propto\epsilon ^3e^{- {\rm const.}\;\epsilon}$.

\item
\underline{$|\kappa|>|\tau|$, $\nu=0$, $\sigma ^2=\kappa ^2-\tau^2$}
$\rightarrow$ {\em the spatially projected worldlines are catenaries,
curves of the type $x=\kappa \cosh (y/\tau)$}.

\bigskip

In general, SVES-5 cannot be found analytically. It is an intermediate
case, which for $\tau/\sigma\rightarrow 0$ tends to SVES-2, whereas
for $\tau/\sigma\rightarrow\infty$ tends toward SVES-4
\begin{equation}
S_2(\epsilon _{\kappa})
\stackrel{0\leftarrow \tau/\sigma}{\longleftarrow}
S_5(\epsilon _{\sigma})\stackrel{\tau/\sigma\rightarrow \infty}
{\longrightarrow}S_4(\epsilon _{\kappa}).
\end{equation}

\item
\underline{$\nu\neq 0$, $\kappa$ and $\tau$ arbitrary}
$\rightarrow$ {\em rotating worldlines uniformly accelerated normal to
their plane of rotation}.

\bigskip

SVES-6 forms a three-parameter set of curves. The corresponding
trajectories are a superposition of the constant linearly accelerated
motion and uniform circular motion. SVES-6 has not been calculated by
Letaw, not even numerically.
\end{enumerate}

Thus, only the hyperbolic worldlines, having just one nonzero
curvature invariant, allow for a Planckian SVES. Only that case allows
for a strictly one-to-one mapping between the curvature invariant
$\kappa$ and the `thermodynamic' temperature in the celebrated form
$T_{U}=\kappa /2\pi$. The vacuum field noise of semicubical parabolas
can be fitted by a Wien-type spectrum, the radiometric parameter then
corresponding to both curvature and torsion.  The other stationary
cases, being nonanalytic, lead to the approximate determination of the
curvature invariants defining locally the classical worldline on which
the relativistic quantum detector moves.

One very general and important statement regarding the {\em universal}
nature of the kinematical Frenet--Serret parameters occurring in
various important quantum field model problems can be formulated as
follows:

\begin{quote}
{\em There exist accelerating classical trajectories (worldlines) on
which moving ideal (two-level) quantum systems can detect the scalar
vacuum environment as a stationary quantum field vacuum noise with a
spectrum directly related to the curvature invariants of the
worldline, thus allowing for a radiometric interpretation of those
invariants}.
\end{quote}

According to these results, it seems more appropriate to replace the
thermal interpretation of Unruh by the radiometric interpretation of
the Frenet--Serret invariants.  The latter is more general and
describes in a more precise way the physical situation to which the
Unruh effect refers, that of a {\em quantum} particle moving along a
{\em classical} relativistic trajectory.

\subsection{Explicit formulae for the spectra}

One can calculate the spectrum of vacuum field noise by means of the
following general formula
\begin{equation}
S_{j}(E)=\left|
\frac{E^2}{4\pi ^3}\int _{-\infty}^{\infty}\frac{e^{-iEs}\; \d s}
{
[x_{\mu}(s)-x_{\mu}(0)]\;[x^{\mu}(s)-x^{\mu}(0)]
}
\right|=
\frac{E^2}{4\pi ^3}|{\rm I}_{j}|,\quad j=1...6,   
\end{equation}
where $x^{\mu}(s)$ is an arbitrary point on the worldline and
$x^{\mu}(0)$ is the initial point. The signature of the Minkowski
metric is $\eta _{\mu \nu}=(1,-1,-1,-1)$.  I confirm Letaw's results
by sketching the calculation of the integrals ${\rm I}_{j}$ for the
six stationary cases.  Simple details that have been skipped by Letaw
can be found here.

\begin{enumerate}
\item {\bf The recta}.
The worldline is $x^{\mu}(s)=(s,0,0,0)$; with initial condition
$x^{\mu}(0)=(0,0,0,0)$. The integral is
\begin{equation}
{\rm I}_1= \int _{-\infty}^{+\infty} \frac{e^{-iEs}}{s^2}\; \d s.
\end{equation}
It can be evaluated by using Cauchy's residue theorem plus the
expansion $e^{-iEs}=(1-iEs+...)$.  The value of the integral is $\pi
i(-iE)=\pi E$, and therefore one gets the cubic spectrum.  This
inertial zero-point cubic spectrum will appear in all the other five
stationary spectra as an additive background and therefore one may
take into account only the non-cubic contributions as a measure of
non-inertial vacuum effects.

\item {\bf The hyperbola}.
The worldline is $x^{\mu}(s)=\kappa ^{-1}(\sinh \kappa s,\cosh \kappa
s,0,0)$; with the initial condition $x^{\mu}(0)=\kappa
^{-1}(0,1,0,0)$. Now the integral is
\begin{equation}
{\rm I}_2=
\kappa\int _{-\infty}^{+\infty}
\frac{e^{-i\epsilon _{\kappa}u}}{2(\cosh u -1)}\;\d u.
\end{equation}
Writing $e^{-i\epsilon _{\kappa}u}=\cos\epsilon _{\kappa} u -
i\sin\epsilon _{\kappa} u$, one makes use of formula 3.983.3 at page
505 in the fourth edition of the {\em Table} of Gradshteyn and Ryzhik
(GR) to get
\begin{equation}
\int _{0}^{+\infty}\frac{\cos(ax)}{\cosh x -1}dx= -\pi a\coth(\pi a).
\end{equation}
The sine integral can be evaluated using Cauchy's theorem
\begin{equation}
-i\int _{-\infty}^{+\infty}
\frac{\sin(\epsilon _{\kappa}x)dx}{\cosh x -1}=
\pi \epsilon _{\kappa}.
\end{equation}
Thus
\begin{eqnarray}
{\rm I}_2 
&=& 
-\pi\epsilon _{\kappa}\coth(\pi\epsilon _{\kappa})
+\frac{\pi \epsilon _{\kappa}}{2}
\\
&=&
\pi\epsilon _{\kappa}\; 
[1-\coth(\pi\epsilon _{\kappa})]-\frac{\pi \epsilon _{\kappa}}{2}
\\
&=&
-2\pi\epsilon _{\kappa}\; \frac{1}{e^{2\pi \epsilon _{\kappa}}-1}
-\frac{\pi\epsilon _{\kappa}}{2},
\end{eqnarray}
where the first term leads to the Planckian spectrum and the latter to
the cubic zero-point contribution.

\item {\bf The helix}.
The worldline is $x^{\mu}(s)=\rho ^{-2}(\tau\rho s,\kappa\cos\rho
s,\kappa\sin \rho s,0)$; with initial condition $x^{\mu}(0)=\rho
^{-2}(0,\kappa,0,0)$.  The integral reads
\begin{equation}
{\rm I}_3=\rho \int _{-\infty}^{+\infty}
\frac{e^{-i\epsilon _{\rho}u}}
{2\frac{\kappa ^2}{\rho ^2}(\cos u-1)+\frac{\tau ^2}{\rho ^2}u^2} 
\;\d u,
\end{equation}
where 
\begin{equation}
\frac{\tau ^2}{\rho ^2}-\frac{\kappa ^2}{\rho ^2}=1.
\end{equation}
According to Letaw this integral is non-analytic and indeed I was not
able to find any helpful formula in the GR {\em Table}.

\item {\bf The semicubical parabola}.

The worldline is $x^{\mu}(s)=(s+\frac{1}{6}\kappa ^2s^3,
\frac{1}{2}\kappa s^2, \frac{1}{6}\kappa ^2s^3,0)$; with initial
condition $x^{\mu}(0)=(0,0,0,0)$. The integral reads
\begin{equation}
{\rm I}_4 
=
\kappa\int _{-\infty}^{+\infty}\frac{e^{-i\epsilon
_{\kappa}u}\; \d u}{u^2(1 +\frac{1}{12}u^2)}
=
\kappa{\rm I}_1-\kappa\int _{-\infty}^{+\infty}
\frac{e^{-i\epsilon _{\kappa}u}\;\d u}{(12+u^2)}.
\end{equation}
Of interest is only the second integral that can be found in the
GR {\em Table} at page 359
\begin{equation}
\int _{-\infty}^{+\infty}\frac{e^{-ipx}\; \d x}{a^2+x^2}=
\frac{\pi}{|a|}\; e^{-|ap|},
\end{equation}
for $a>0$ and $p$ real. Thus one gets
\begin{equation}
\int _{-\infty}^{+\infty}
\frac{e^{-i\epsilon _{\kappa}u}\; \d u}{(12+u^2)}=
\frac{\pi}{\sqrt{12}}\; e^{-\sqrt{12}\epsilon _{\kappa}}.
\end{equation}
The final result is
\begin{equation}
S_4=
\frac{\kappa ^4\epsilon _{\kappa}^{2}}{4\pi ^2\sqrt{12}}
\; e^{-\sqrt{12}\epsilon _{\kappa}}.
\end{equation}

Interestingly, for a horizontal storage ring (guiding magnetic field
in the vertical $z$ direction) the orbit in the moving frame can be
approximated for laboratory times such that $\gamma \omega
_{0}|t|=O(1)$ by the following semicubical parabola
\begin{equation}
y'\approx \left({R_0 \over 2\gamma ^2}\right)\; 
\left({6\gamma ^2|x'|\over R_0}\right)^{2/3},
\end{equation}
where $R_0$ is the instantaneous radius of curvature of a particle's
arbitrary trajectory.\footnote{See Fig. 2 and Eq. 2 in \cite{J}.}

\item {\bf The catenary}.

The worldline is $x^{\mu}(s)=\sigma ^{-2}(\kappa\sinh\sigma
s,\kappa\cosh\sigma s, \tau\sigma s,0)$; with initial condition
$x^{\mu}(0)=\sigma ^{-2} (0,\kappa,0,0)$. The integral is of the type
\begin{equation}
{\rm I}_5=\sigma
\int _{-\infty}^{+\infty}\frac{e^{-i\epsilon _{\sigma}u} \; \d u}
{2\frac{\kappa ^2}{\sigma ^2}({\rm cosh} u-1)-
\frac{\tau ^2}{\sigma ^2}u^2}, 
\end{equation}
where $\frac{\tau ^2}{\sigma ^2}-\frac{\kappa ^2}{\sigma ^2}=-1$.
This integral turns into $I_2$ and $I_4$ in the limits mentioned in
the text, respectively, but again there is no helpful formula in the
GR {\em Table}, and thus ${\rm I}_5$ appears to be non-analytic.

\item {\bf The helicoid (helix of variable pitch)}.

This is the most general case.  The worldline is
\begin{eqnarray}
x^{\mu}(s)&=&\Bigg(
\frac{\Delta}{RR_{+}} \sinh (R_{+}s), 
\frac{\Delta}{RR_{+}} \cosh(R_{+}s), 
\nonumber
\\
&&
\qquad
\frac{\kappa\tau}{R\Delta R_{-}} \cos(R_{-}s),
\frac{\kappa \tau}{R\Delta R_{-}} \sin(R_{-}s)
\Bigg);
\end{eqnarray}
while the initial condition reads
$x^{\mu}(0)=(0,\frac{\Delta}{RR_{+}},\frac{\kappa \tau}{R\Delta
R_{-}},0)$.  I have defined: 
\begin{eqnarray}
\Delta ^2&=&\frac{1}{2}(R^2+\kappa ^2 +\tau ^2 +\nu^2); 
\\
R^4&=&(\kappa ^2+\tau ^2 +\nu ^2)^2-4\kappa ^2 \tau ^2;
\\
R_{+}^2&=&\frac{1}{2}(R^2+\kappa ^2 -\tau ^2 -\nu ^2);
\\
R_{-}^2&=&\frac{1}{2}(R^2-\kappa ^2 +\tau ^2 +\nu ^2).
\end{eqnarray}

\noindent
The following integral is obtained:
\begin{equation}
{\rm I}_6=
R\int _{-\infty}^{+\infty}
\frac{e^{-i\epsilon _Ru}du}{2(\frac{\Delta}{R_{+}})^2
[{\cosh} (\frac{R_{+}}{R}u)
-1]
+2(\frac{\kappa \tau}{\Delta R_{-}})^2[{\cos}(\frac{R_{-}}{R}u)-1]}.
\end{equation}
This is the most complicated non-analytic stationary case, with no
helpful formula in the GR {\em Table}.
\end{enumerate}

\subsection{Non-stationary vacuum field noise}

Non-stationary vacuum field noise has a time-dependent spectral
content requiring joint time and frequency information, \ie, one needs
generalizations of the power spectrum concept. One can think of (i)
tomographical processing and/or (ii) wavelet transforms.  For
instance, the recently proposed non-commutative tomography (NCT)
transform $M(s;\mu,\nu)$~\cite{MVM} seems to be an attractive way of
processing non-stationary signals.  In the definition of $M$, $s$ is
just an arbitrary curve in the non-commutative time-frequency plane,
while $\mu$ and $\nu$ are parameters characterizing the curve. The
most simple examples are the axes $s=\mu t+\nu \omega$, where $\mu$
and $\nu$ are linear combination parameters.  The non-commutative
tomography transform is related to the Wigner--Ville
quasi-distribution $W(t,\omega)$ by an invertible transformation and
has the following useful properties
\begin{eqnarray}
M(t;1,0)&=&|f(t)|^2,
\\
M(\omega ;0,1)&=&|f(\omega)|^2,
\end{eqnarray}
where $f$ is the analytic signal which is simulated by $M$.
Furthermore, employing $M$ leads to an enhanced detection of the
presence of signals in noise which has a small signal-to-noise ratio.
The latter property may be very useful in detecting VFNs, which are
very small `signals' with respect to more common noise sources.

On the other hand, since in the quantum detector method the vacuum
autocorrelation functions are the essential physical quantities, and
since according to various fluctuation-dissipation theorems they are
related to the linear (equilibrium) response functions to an initial
condition/vacuum, the fluctuation-dissipation approach has been
developed and promoted by Sciama and collaborators~\cite{Sc}.  In
principle, the generalization of the fluctuation-dissipation theorem
for some classes of out of equilibrium relaxational systems, such as
glasses, looks also promising for the case of non-stationary vacuum
noise.  One can use a so-called two-time fluctuation-dissipation ratio
$X(t,t')$ and write a modified fluctuation-dissipation relationship
\cite{glass}
\begin{equation}
T_{\rm eff}(t,t') \; R(t,t')=X(t,t') \; 
\frac{\partial C(t,t')}{\partial t'},
\end{equation}
where $R$ is the response function and $C$ the autocorrelation
function.  The fluctuation-dissipation ratio is employed to perform
the separation of scales. Moreover, $T_{\rm eff}$ are
timescale-dependent quantities, making them promising for relativistic
VFNs, which correspond naturally to out of equilibrium conditions.

\section{Circular electromagnetic vacuum noise} 

\subsection{Introduction}

The circular electromagnetic vacuum noise, which in principle is more
promising experimentally,\footnote{%
Rogers \cite{rog} proposed to study the motion of a single electron in
a Penning trap (geonium) to detect the circular electromagnetic vacuum
noise.  For two-level atoms in circular motion the reader is referred
to Audretsch {\em et al}~\cite{Aud}, whereas in the analog style
approach Calogeracos and Volovik \cite{CV} considered the
quasiparticle radiation from objects rotating in superfluid vacuum.}
has been first discussed for specific purposes by Candelas and
Deutsch, and by Bell and Leinaas.  However, here we will pay more
attention to the approach of Hacyan and Sarmiento (HS)~\cite{HS} who
in 1989 introduced a clear-cut and general method for calculating the
main electromagnetic vacuum spectral quantities and applied it to the
basic cases of {\em linear acceleration} and {\em uniform rotation}.
In the latter case, they obtained a nonzero energy flux in the
direction of motion of the detector. It was this result that prompted
Mane~\cite{M} to suggest a connection with the synchrotron radiation.

In principle, the circular vacuum noise power spectrum $S_{\rm c}$
could be calculated via the residue theorem, but the equation for the
zeros of the denominator $x^2=v^2\sin ^2 x$ (see below) is not
analytically solvable. Nevertheless, for $v \geq 0.85$ one can expand
the sine to find the zero with the smallest imaginary part, besides
$x=0$~\cite{Ur}.

\subsection{The Hacyan--Sarmiento approach}
 
Starting with the expression for the electromagnetic energy-momentum
tensor
\begin{equation} 
\label{hs1}
T_{\mu \nu}=\frac{1}{16\pi}\left( 
4F_{\mu \alpha}\;F_{\nu}^{\alpha}+ 
\eta _{\mu \nu}\;F_{\lambda \beta}\;F^{\lambda \beta}
\right).
\end{equation}
Hacyan--Sarmiento define the electromagnetic two-point Wightman
functions as follows
\begin{equation} 
\label{hs2}
D_{\mu \nu}^{+}(x,x')\equiv 
\frac{1}{4}
\left( 
4F^\alpha{}_{(\mu}(x) \;F_{\nu )\alpha}(x')+
\eta _{\mu \nu}\; F_{\lambda \beta}(x)\; F^{\lambda \beta}(x')\right);
\end{equation}
\begin{equation} 
\label{hs3}
D_{\mu \nu}^{-}(x,x')\equiv D_{\mu \nu}^{+}(x',x).
\end{equation}
This may be viewed as a variant of the ``point-splitting'' approach
advocated by DeWitt.  Moreover, because of the properties
\begin{equation} 
\label{hs4}
\eta^{\mu \nu}\;D_{\mu \nu}^{\pm}=0, 
\qquad D_{\mu \nu}^{\pm}=D_{\nu \mu}^{\pm}, 
\qquad
\partial _{\nu}D_{\mu}^{\pm \nu}=0,
\end{equation}
the electromagnetic Wightman functions can be expressed in terms of
the scalar Wightman functions as follows
\begin{equation} 
\label{hs5}
D_{\mu \nu}^{\pm}(x,x')=c\; \partial_\mu \; \partial_\nu \;D^{\pm}(x,x'),
\end{equation}
where $c$ is in general a real constant depending on the case under
study.  This shows that from the standpoint of their vacuum
fluctuations the scalar and the electromagnetic fields are not so
different.

Now introduce sum and difference variables
\begin{equation}
s = {t + t'\over2}; \qquad \sigma = {t-t'\over2}.
\end{equation}
Using the Fourier transforms of the Wightman functions
\begin{equation} 
\label{ksy1}
\tilde{D}^{\pm}(\omega , s)= 
\int _{-\infty}^{\infty} \d\sigma \; 
e^{-i\omega \sigma}\; D^{\pm}(s,\sigma),
\end{equation}
where $\omega$ is the frequency of zero-point fields, the {\em
particle number density of the vacuum seen by the moving detector} and
the {\em spectral vacuum energy density per mode} are given by
\begin{equation} \label{ksy2}
n(\omega , s)=\frac{1}{(2\pi)^2 \omega}
\left[
\tilde{D}^{+}(\omega , s )-
\tilde{D}^{-}(\omega , s )
\right],
\end{equation}
\begin{equation} \label{ksy3}
\frac{\d e}{\d\omega}=\frac{\omega ^2}{\pi}
\left[
\tilde{D}^{+}(\omega , s )+
\tilde{D}^{-}(\omega , s )
\right].
\end{equation}
The most important application of these results is to a uniformly
rotating detector whose proper time is $s$ and angular speed is
$\omega _{0}$ in motion along the circular world line
\begin{equation} 
\label{hswl}
x^{\alpha}(s)=
(\gamma s, R_{0}\cos (\omega _{0}s),
R_{0}\sin (\omega _{0}s), 0),
\end{equation}
where $R_{0}$ is the rotation radius in the inertial frame, 
$\gamma =(1-v^2)^{-1/2}$, and $v={\omega _0 R_0}/{\gamma}$.
\footnote{%
This is correct for Galilean electromagnetism and works well at low
velocities and/or in gradient index (lens) media. For full Lorentz
covariant electrodynamics, one should use the Trocheris-Takeno
nonlinear relationship $v=\tanh (\omega R_0/c)$. See, {\emph{e.g.}},
\cite{TT}.}
In this case there are two Killing vectors $k^{\alpha}=(1,0,0,0)$ and
$ m^{\alpha}(s)=(0,-R_{0}\sin(\omega_{0}s), R_{0}
\cos(\omega_{0}s,0)$. Expressing the Wightman functions in terms of
these two Killing vectors, HS calculated the following physically
observable spectral quantities (\ie, those obtained after subtracting
the inertial zero-point field contributions):

\begin{itemize}
\item {\em The spectral energy density}
\begin{equation} \label{hsA}
\frac{\d e}{\d\omega}=
\frac{\gamma ^3}{2\pi ^3 R_{0}^{3}} \;
\frac{\omega ^2+(\gamma v \omega _{0})^2}{\omega ^2} \;
\frac{v^3w^2}{w^2+(2\gamma v)^2} \; 
h_{\gamma}(w),
\end{equation}
\item {\em The spectral flux density}
\begin{equation} \label{hsB}
\frac{\d p}{\d\omega}=
\frac{\gamma ^3}{2\pi ^3 R_{0}^{3}}\;
\frac{\omega ^2+(\gamma v \omega _{0})^2}{\omega ^2}\;
4v^4 \; k_{\gamma}(w),
\end{equation}
\item {\em The spectral stress density}
\begin{equation} \label{hsC}
\frac{\d s}{\d\omega}=
\frac{\gamma ^3}{2\pi ^3 R_{0}^{3}}\;
\frac{\omega ^2+(\gamma v \omega _{0})^2}{\omega ^2}\;
\frac{v^3w^2}{w^2+(2\gamma v)^2}\;
j_{\gamma}(w).
\end{equation}
\end{itemize}

Here $(\omega ^2+(\gamma v \omega _{0})^2)/\omega ^2$ is a
density-of-states factor introduced for convenience and
$h_{\gamma}(w)$, $k_{\gamma}(w)$, and $j_{\gamma}(w)$ are the
following cosine-Fourier transforms
\begin{equation} 
\label{hsD}
h_{\gamma}(w)\equiv   
\int _{0}^{\infty}\left(\frac{N_h(x,v)}
{\gamma ^2[x^2-v^2\sin ^2 x]^3}
-\frac{3}{x^4}+\frac{2\gamma ^2 v^2}{x^2}\right)\cos (wx)\; \d x;
\end{equation}
\begin{equation} 
\label{hsE}
k_{\gamma}(w)\equiv -
\int _{0}^{\infty}\left(\frac{N_k(x,v)}{\gamma ^2[x^2-v^2\sin ^2 x]^3}
-\frac{3}{x^4}-\frac{\gamma ^2}{6x^2}\right)
\cos (wx)\; \d x;
\end{equation}
\begin{equation} 
\label{hsF}
j_{\gamma}(w)\equiv 
\int _{0}^{\infty}\left(\frac{1}{\gamma ^4[x^2-v^2\sin ^2 x]^2}-
\frac{1}{x^4}+\frac{2\gamma ^2 v^2}{3x^2}\right)\cos (wx)\; \d x.
\end{equation}
The numerators $N_h(x,v)$ and $N_k(x,v)$ are given by
\begin{eqnarray}
N_h(x,v)&=&(3+v^2)x^2+(v^2+3v^4)\sin ^2x-8v^2x \sin x;
\\ 
N_k(x,v)&=&x^2+v^2\sin ^2x-(1+v^2)x \sin x. 
\end{eqnarray}
The employed variables are
$w=\frac{2\omega}{\omega _{0}}$ and $x=\frac{\sigma \omega _{0}}{2}$.

Of special interest are the ultra-relativistic and nonrelativistic
limits. In the first case, $\gamma \gg 1$, the quantities
\begin{equation}
H_{\gamma}=\frac{v^3w^2}{w^2+(2\gamma v)^2}\;
h_{\gamma}(w),
\quad
K_{\gamma}=4v^{4}\; k_{\gamma}(w),
\quad 
J_{\gamma}=\frac{v^3w^2}{w^2+(2\gamma v)^2}\; j_{\gamma}(w),
\end{equation}
have the following scaling property
\begin{equation}
X_{k\gamma}(kw)=k^3X_{\gamma}(w),
\end{equation}
where $k$ is an arbitrary constant, and $X=H,K,J$. This is the same
scaling property as that of a Planckian distribution with a
temperature proportional to $\gamma$.


A detailed discussion of the nonrelativistic limit has been provided
by Kim, Soh, and Yee~\cite{KSY}, who used the parameters $v$ and
$\omega _{0}$, and not acceleration and speed as used by Letaw and
Pfautsch for the circular scalar case~\cite{LP}.  They obtained a
series expansion in velocity
\begin{equation} 
\label{ksy6}
\frac{de}{d\omega}=
\frac{\omega ^3}{\pi ^2}
\left[
\frac{\omega _{0}}{\gamma\omega}
\sum _{n=0}^{\infty}
\frac{v^{2n}}{2n+1}\sum _{k=0}^{n}(-1)^{k}
\;
\frac{(n-k-\frac{\omega}{\gamma \omega _0})^{2n+1}}{k!\; (2n-k)!}
\;
H\left(n-k-\frac{\omega}{\gamma \omega _0}\right)\right],
\end{equation}
where $H$ is the usual Heavyside step function. Thus, to a specified
power of the velocity many vacuum harmonics could contribute; making
the energy density spectrum quasi-continuous.

\subsection[Synchrotron radiation as vacuum fluctuations?]{%
Synchrotron radiation as electromagnetic vacuum fluctuations ?} 

In 1991, Mane used the Hacyan--Sarmiento formula for the energy flux
to argue that its time component is related to the synchrotron
radiation.  The Hacyan--Sarmiento Poynting flux is directed along the
Lorentz boost from the laboratory frame to the rest frame of the
observer, which is taken as the $y$ axis. It can be written
\begin{equation} 
\label{mane1}
p_{y}=\frac{1}{1440\pi ^2}\;
\frac{\hbar \gamma ^8 \omega _{0}^4 v}{c^4}
\; (50 -47 \gamma ^{-2}).
\end{equation}
Note that $p_{y}$ is proportional to $\hbar$ and therefore becomes
zero in the classical limit. However, for electrons which couple to
this flux via the fine structure constant $\alpha =e^2/\hbar c$, the
radiation effect looks totally classical.  The recoil induced by the
flux of the vacuum fluctuations on the four-momentum of the particle
per unit proper time is
\begin{equation} 
\label{mane2}
\alpha \, A \, p_{y}\propto \frac{e^2\gamma ^4\omega _{0}^2 v}{c^3},
\end{equation}
where $A\approx R_{0}^{2}c^2/(v^2\gamma ^4)$ is the transverse
interaction area between the electron and the electromagnetic field.
In the laboratory frame, the energy loss of the particle per unit
laboratory time is given by the Larmor formula
\begin{equation} 
\label{mane3}
I=\frac{2}{3}\;\frac{e^2}{c^3}(\gamma ^2 \omega _{0}v)^2.
\end{equation}
This is related to the damping force $\vec{F}$ in the form
$I=\vec{F}\cdot \vec{v}$ and therefore the recoil induced by
synchrotron radiation on the four-momentum of the particle per unit
proper time is again proportional to ${e^2\gamma ^4\omega _{0}^2
v}/{c^3}$ as in Eq.~(\ref{mane2}).  Therefore, the order of magnitude
of the recoil of the particle induced by the $\alpha$ coupling to the
vacuum flux is equal to that derived by the Larmor formula in the
ultra-relativistic limit.

If one goes as far as accepting the idea that synchrotron radiation is
due to noninertial electromagnetic vacuum fluctuations, one should
reproduce in this approach all the many basic features of the
synchrotron radiation that are known from both theory and measurements
at storage rings.  Recall, for example, that the Schwinger spectral
intensity of the magneto-bremsstrahlung in the synchrotron
regime~\cite{S} is proportional to the so-called shape function
\begin{equation} 
\label{sw1}
 W_{\omega}\propto
F\left(\frac{\omega}{\omega _m}\right),
\end{equation}
where $\omega _m$ is given in terms of the cyclotron radian frequency
$\omega _c$ as $\omega _m =\omega _c \gamma ^3$, and the shape
function $F$ is given by
\begin{equation}
F(\zeta) =
\frac{9\sqrt{3}}{8\pi}\;\zeta \;
\int _{\zeta} ^{\infty} K_{5/3}(z)\; \d z,
\end{equation} 
where $K$ is the MacDonald (modified Bessel) function of the quoted
fractional order. The small and large asymptotic limits of the
synchrotron shape function are as follows
\begin{equation} 
\label{sw2}
F(\zeta \ll 1)\approx 1.33\;\zeta ^{1/3},
\end{equation}
and
\begin{equation} \label{sw3}
F(\zeta \gg 1)\approx 0.78\;\zeta ^{1/2}\;e^{-\zeta},
\end{equation}
with a maximum (amount of radiation) to be found at the frequency
$\omega_{m}/3$. An examination of the Hacyan--Sarmiento asymptotic
limits shows that there are clear differences between the
Hacyan--Sarmiento and synchrotron energy density spectrum. Neither of
the two limits coincide, neither the Hacyan--Sarmiento spectrum
divided in two equal parts by its peak frequency as in the case of
synchrotron radiation. Moreover, the well-defined polarization state
of synchrotron radiation that can be calculated in closed form in
terms of the squares of Bessel $K_{1/3}$ and $K_{2/3}$ functions in
electrodynamics would prove really difficult to obtain in the vacuum
approach.  Finally, an expansion in velocity powers of the synchrotron
radiation does not coincide with that in Eq.~(\ref{ksy6}).  In the
opinion of the author, the circular electromagnetic vacuum noise
should be considered as only a radiation signal embedded in the
synchrotron radiation background.

\subsection[Electron beam polarization at storage rings]{%
Electron beam polarization at storage rings: 
\\ Spin flip synchrotron radiation versus circular electromagnetic 
vacuum noise}

Electromagnetic circular vacuum noise is interesting not only because
of the Hacyan--Sarmiento results and Mane's suggestion but also as
being responsible, according to Bell and Leinaas~\cite{BL}, for the
electron depolarization at storage rings. This famous proposal was put
under intense focus in 1998 at the Monterey conference organized by
Pisin Chen, where one of the most authoritative contrarians, Professor
J.D. Jackson declared~\cite{J2}:
\begin{quote}
{\emph{ Avoid the indiscriminate appeal to Unruh in order to
``understand'' something amenable to a simpler explanation.}}
\end{quote}
The following is a brief introduction to this problem. It has been
included here as an illustration of the confrontation of theoretical
ideas with the experimental market; a confrontation eagerly awaited
for the attractive analog proposals presented in this book.

The Bell-Leinaas proposal relies on the spin degree of freedom of the
electron in an external magnetic field $B_0$ along the $z$ axis. The
spin may be thought to have two (quasi)stationary states corresponding
to $\sigma _{z}=\pm 1$, with an energy splitting $\Delta =2|\mu|\;
|B_{0}|$.  This approximation is valid when a second term in the
effective spin-field interaction Hamiltonian, due to the so-called
Thomas precession, is not included.  Thus, the electron looks in a
first approximation like an Unruh--DeWitt detector.  The transitions
between the two spin states induced by the radiation field are then
written in terms of first-order time-dependent perturbation theory,
and a thermal ratio is obtained as if produced by the equilibrium
ratio of populations of the upper and lower levels. The effect of the
Thomas precession term in the effective Hamiltonian does not alter the
shape of the polarization curve, and only shifts it horizontally when
plotted as a function of the magnetic moment.

However, there is a simple quantum electrodynamical explanation of the
polarization effect at storage rings in terms of the so called
spin-flip synchrotron radiation that has been proposed by Sokolov and
Ternov in 1963 \cite{ST}. The spin-flip radiated power is very small
with respect to the ordinary synchrotron radiation, becoming of the
same order only at $\gamma _{sf}=(mcR_0/\hbar)^{1/2}$, which for a
common storage ring is around $\gamma \approx 6 \times 10^{6}$. This
is more than two orders of magnitude higher than the actual $\gamma <
10^{4}$ of electrons in current storage rings leading to a spin-flip
radiated power representing only $10^{-11}$ of the usual (non
spin-flip) synchrotron emitted power. It is only because the spin flip
accumulates over a time scale of tens of minutes to a few hours that
one gets the observed asymptotic polarization
$P_{lim}=8/5\sqrt{3}=0.924$.

It is either this non-stationarity of the spin-flip synchrotron
radiation, or the fact that the orbiting electrons are actually more
complicated interacting systems than simple Unruh--DeWitt detectors,
that lead only to frequency-dependent effective temperatures, which in
the opinion of accelerator people are not useful parameters.

The description of radiative polarization in terms of spin levels came
under the scrutiny of Professor Jackson long ago~\cite{J}. He showed
that the spacing between orbital levels is very small compared to the
magnetic dipole M1 transition energy, and therefore the M1 transition
will involve some changes in the orbital quantum number. In 1973,
Derbenev and Kondratenko~\cite{DK} obtained, in a quasi-classical
approach, a formula for the equilibrium polarization in which
spin-orbit effects are included through a spin-orbit coupling
function.  Their formula is considered to be the standard result for
the transverse polarization at storage rings. According to the
Derbenev--Kondratenko formula, for the range $0<g<1.2$ one of the
levels is preferentially populated with respect to the other one.
This effect cannot be reproduced in the Bell--Leinaas approach without
resorting to time-dependent couplings and frequency-dependent
`temperatures`~\cite{Ur}.  In 1987, Bell and Leinaas published a more
detailed analysis of their proposal, in which (still assuming a
thermal spectrum of the spin excitations) they took into account the
fluctuations in the orbital motion.  They obtained a polarization
formula rather close to the standard one with some differences only
close to a narrow depolarizing resonance. The claim is that when
passing through the resonance the polarization falls from 92\% to
-17\% followed by an increase to 99\% before settling again to 92\%.

Thus, the confirmation of their calculation, and of the thermal vacuum
bath, as opposed to the Sokolov--Ternov limiting polarization would
require precise experimental measurements of a transient passing
through a depolarizing resonance, an experiment that is still to be
performed.

\section{Unruh effect {\emph{versus}} anomalous Doppler effect}
\label{S:ade}

The concept of the anomalous Doppler effect (ADE) was introduced in
classical electrodynamics by Frank in 1942~\cite{Frank}.  ADE refers
to the waves emitted {\emph{within}} the Cherenkov cone by a
``superluminal'' oscillator moving in a refractive medium. (Frank's
example is an electric dipole harmonically oscillating at angular
frequency $\Omega$.) By definition, these waves exhibit an anomalous
Doppler shift in the sense that their frequencies (with respect to
$\Omega$) have a negative Doppler directivity factor $D$ (see below)
and are given by $\omega_{ADE}=\Omega /D$. In the quantum version of
this phenomenon, as discussed by Frolov and Ginzburg, one uses the
energy-momentum conservation law for massless Bose radiation from a
superluminal two-level detector to get the energy formula (\ref{ADE1})
below.

\begin{figure}[htb]
\vbox{
\vskip 20 pt

\centerline{\epsfxsize=5.0in\epsffile{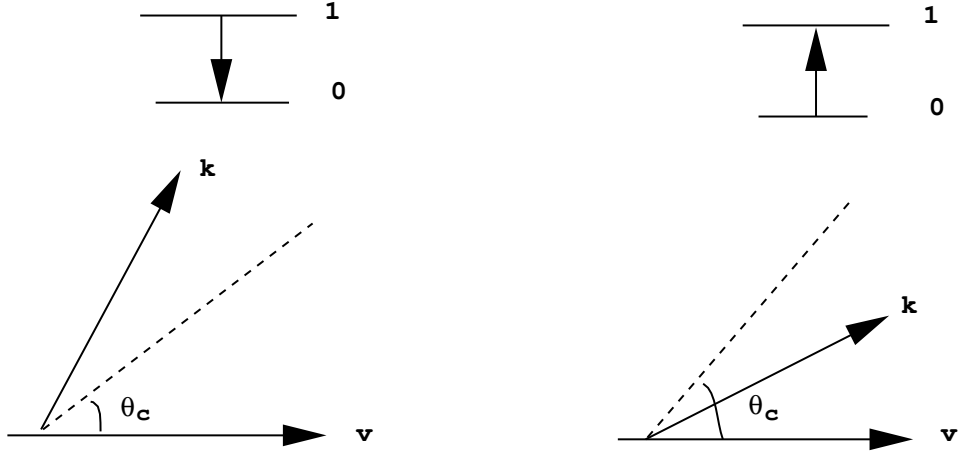}}

\caption[Normal and anomalous Doppler effects]{%
\label{F:ade}
{\sl The normal and anomalous Doppler effects and the corresponding
transitions.
\smallskip}
}
}
\end{figure}

As we have already mentioned in Section (\ref{S:udw}), when studied
with the detector method the Unruh effect for a detector with internal
degrees of freedom is in some ways very similar to this anomalous
Doppler effect (ADE), since in both cases the quantum detector is
radiating `photons' while passing into the upper level and not on the
lower one~\cite{FG} (see Fig.\ref{F:ade}).

This is refrained in the well-known conclusion of Unruh and
Wald~\cite{wal} when they considered the uniformly accelerated quantum
detector looked upon from the inertial reference frame:
\begin{quote}
{\emph{When the observer places himself in an inertial reference frame
then he is able to observe both the excited quantum detector
(furnishing at the same time energy to it) and the `photons'. By
writing down the energy-momentum conservation law he will be inclined
to say that the `photons' are emitted precisely when the detector is
excited.}}
\end{quote}
Neglecting recoil, absorption, and dispersion (a completely ideal
case) the elementary radiation events for a two-level detector with
the change of the detector proper energy denoted by $\delta\epsilon$
are classified according to the photon energy formula~\cite{FG}
\begin{equation} 
\label{ADE1}
\hbar\omega ={ -\frac{\delta\epsilon}{D\gamma}}
\end{equation}
where $\gamma$ is the relativistic velocity factor ($\gamma> 1$) and
D is the Doppler directivity factor
\begin{equation} 
\label{ADE2}
D = 1 - \left({vn\over c}\right) \cos\theta.
\end{equation}
The discussion of signs in Eq.~(\ref{ADE1}) implies 3 cases as follows:
\begin{enumerate}
\item
D$ >$ 0 for normal Doppler effect (NDE, $\delta\epsilon<0$, 
  $\theta >\theta _c$)
\item
D = 0 for Cherenkov effect (CE, $\delta\epsilon=0$, undetermined case,
$\theta =\theta _c$)
\item
D$ <$ 0 for anomalous Doppler effect (ADE, $\delta\epsilon>0$,
$\theta <\theta _c$).
\end{enumerate}
Consequently, for a quantum system endowed with internal degrees of
freedom the stationary population of levels is determined by the
probability of radiation in the ADE and NDE regions. The possibility
of inducing population inversion by means of the ADE has been
extensively discussed in the literature since a long time
ago~\cite{unpub}.

Bolotovsky and Bykov~\cite{bob} have studied the space-time properties
of ADE in the simple case of a superluminal dipole ($v>c/n$)
propagating in uniform rectilinear motion in a non-dispersive
medium. These authors claim a positive theoretical result with regard
to the separate observation of the ADE phenomenon for this case. It is
not, unfortunately, a realistic case and requires a special equation
of motion of the dipole.  Theoretical and experimental investigations of
the possible manifestation of ADE in dispersive and/or lens media is
an important task for the future.

A direct experimental evidence of ADE would be highly valuable as
being equivalent to a test of the Unruh effect.  The acoustic ADE is
another challenge for the future \cite{Marvin}.

\section{Summary}

As is the case for all claimed quantum vacuum effects (including the
mechanical Casimir effect), the stationary radiative spectra surveyed
in this chapter can be attributed equally well to radiation reaction
fields, \ie, to solutions of the inhomogeneous Klein-Gordon or Maxwell
equations evaluated at the source~\cite{bd}. However, the main point
we want to emphasize is different, namely: These radiation patterns
(if they really exist --- remember the calculations are all performed
in first-order perturbation theory, or in order $\alpha$ for
electrodynamic radiation reaction fields) can be used to extract
radiometric information related to the Frenet--Serret geometric
invariants of the trajectories of the relativistic corpuscules.

Finally, we emphasise that over the last few years the physics
community has become aware of many interesting similarities/analogies
between the Hawking/Unruh effects and shock-type effects in material
media. A very promising line of research could be the study of the
Cherenkov effect, and the associated anomalous and normal Doppler
effects of relativistic dipoles propagating in strongly dispersive
substances.  Potentially realistic laboratory configurations for
examining these effects are, for example, Cherenkov-type experiments
with bunches of electric dipoles (polarization pulses) created by
femtosecond optical pulses in electro-optic materials~\cite{eo}. In
addition, radiation from vortices in two-dimensional annular Josephson
junctions~\cite{kysv}, or even from other `relativistic' defects in
condensed-matter physics should be taken into account from the
perspective of this book.

\bigskip
{\bf Acknowledgements:} The author would like to thank the organizers
of the workshop for inviting him to this interesting event.




\begin{thebibliography} {99}

\bibitem{T} For review, see S. Takagi,
``Vacuum noise and stress induced by uniform acceleration", 
Prog. Theor. Phys. Suppl. {\bf 88}, 1  (1986).

\bibitem{L}  
J.R. Letaw, 
``Stationary world lines and the vacuum excitation of 
noninertial detectors",
 Phys. Rev. D {\bf 23}, 1709 (1981).

\bibitem{FG}
V.P. Frolov and V.L. Ginzburg, 
``Excitation and radiation of an accelerated detector and anomalous 
Doppler effect", 
Phys. Lett. A {\bf 116}, 423 (1986).


\bibitem{un}
W. Unruh, 
``Radiation reaction fields for an accelerated dipole for scalar and 
electromagnetic radiation",
Phys. Rev. A {\bf 59}, 131 (1999).

\bibitem{Ho}
E. Honig, E.L. Schucking, and C.V. Vishveshwara,
``Motion of charged particles in homogeneous electromagnetic fields",
J. Math. Phys. {\bf 15}, 774 (1974). 
\\
For higher-dimensional generalization,
see B.R. Iyer and C.V. Vishveshwara, 
``The Frenet--Serret formalism and black holes in higher dimensions",
Class. Quant. Grav. {\bf 5}, 961 (1988).

\bibitem{MVM} 
V.I. Man'ko and R. Vilela Mendes, 
``Noncommutative time-frequency tomography", 
Phys. Lett. A {\bf 263}, 53 (1999) [physics/9712022].

\bibitem{Sc}
P. Candelas and D.W. Sciama, 
``Irreversible thermodynamics for black holes",
Phys. Rev. Lett. {\bf 38}, 1372 (1977).


\bibitem{glass}
 L.F. Cugliandolo, J. Kurchan and L. Peliti, 
``Energy flow, partial equilibration and effective temperatures 
in systems with slow dynamics",
Phys. Rev. {\bf E 55}, 3898 (1997) [cond-mat/9611044]; 
\\
R. Exartier and L. Peliti, 
``Measuring effective temperatures in nonequilibrium systems", 
Eur. Phys. J. {\bf B 16}, 119 (2000), [cond-mat/9910412].


\bibitem{rog}
J. Rogers, 
``Detector for the temperature-like effect of acceleration",
Phys. Rev. Lett. {\bf 61}, 2113 (1988).

\bibitem{Aud}
J. Audretsch, R. M\"uller, and M. Holzmann, 
``Generalized Unruh effect and Lamb shift on arbitrary 
stationary trajectories",
Class. Quant. Grav. {\bf 12}, 2927 (1995).

\bibitem{CV}
A. Calogeracos and G.E. Volovik, 
``Rotational quantum friction in superfluids: Radiation from 
object rotating in superfluid vacuum",
Pis'ma Zh. Eksp. Teor. Fiz. {\bf 69}, 257 (1999).


\bibitem{HS} 
S. Hacyan and A. Sarmiento, 
``Vacuum stress-energy tensor of the electromagnetic  field in 
rotating frames", 
Phys. Rev. D {\bf 40}, 2641 (1989); 
Phys. Lett. B {\bf 179}, 287 (1986).

\bibitem{M}
S.R. Mane, ``Comment on HS",
Phys. Rev. D {\bf 43}, 3578 (1991);
\\
See also, T. Hirayama and T. Hara, 
``A calculation on the self-field of a point charge and the 
Unruh effect", 
Prog. Theor. Phys. {\bf 103}, 907 (2000) [gr-qc/9910111].

\bibitem{Ur}
W.G. Unruh, 
``Acceleration radiation for orbiting electrons", 
Phys. Rept. {\bf 307}, 163 (1998) [hep-th/9804158].

\bibitem{TT}
S. Kichenassamy and R.A. Krikorian, 
``Note on Maxwell's equations in relativistically rotating frames",
J. Math. Phys. {\bf 35}, 5726 (1994);
\\
R.D.M. De Paola and N.F. Svaiter,
``A rotating vacuum and a quantum version of Newton's bucket experiment",
Class. Quant. Grav. {\bf 18}, 1799 (2001) [gr-qc/0009058].


\bibitem{KSY} 
S.K. Kim, K.S. Soh, and J.H. Yee, 
``Zero-point field in a circular-motion frame",
Phys. Rev. D {\bf 35}, 557 (1987).

\bibitem{LP} 
J.R. Letaw and J.D. Pfautsch, 
``Quantized scalar field in rotating coordinates", 
Phys. Rev. D {\bf 22}, 1345 (1980). 

\bibitem{S}
J. Schwinger, 
``On the classical radiation of accelerated electrons",
Phys. Rev. {\bf 75}, 1912 (1949).

\bibitem{BL}
J.S. Bell and J.M. Leinaas, 
``Electrons as accelerated thermometers",
Nucl. Phys. B {\bf 212}, 131 (1983); 
``The Unruh effect and quantum fluctuations of electrons in 
storage rings", 
Nucl. Phys. B {\bf 284}, 488 (1987).

\bibitem{J2}
J.D. Jackson, 
``On effective temperatures and electron spin 
polarization in storage rings", 
in {\em Quantum aspects of beam physics}, ed. Pisin Chen,
World Scientific (1999) pp. 622-625 [physics/9901038].

\bibitem{ST}
A.A. Sokolov and I.M. Ternov, 
``On polarization and spin effects in the theory of synchrotron radiation'',
Dokl. Akad. Nauk {\bf 153}, 1052 (1964) 
[Sov. Phys. Dokl. {\bf 8}, 1203 (1964)].

\bibitem{J}
J.D. Jackson, 
``On understanding spin-flip synchrotron radiation and the transverse 
polarization of electrons in storage rings", 
Rev. Mod. Phys. {\bf 48}, 417 (1976). 
\\
For a recent review see, D.P. Barber,
``Electron and proton spin polarization in storage rings 
--- an introduction", 
in {\em Quantum aspects of beam physics}, ed. Pisin Chen,
World Scientific (1999) pp. 67-90 [physics/9901038].

\bibitem{DK}
Ya. S. Derbenev and A.M. Kondratenko, 
``Polarization kinetics of particles in storage rings", 
Zh. Eksp. Teor. Fiz.  {\bf 64}, 1918 (1973) 
[Sov. Phys. JETP {\bf 37}, 968 (1973)].

\bibitem{wal}
W.G. Unruh and R.M. Wald, 
``What happens when an accelerating observer detects a Rindler particle",
Phys. Rev. D {\bf 29}, 1047 (1984).

\bibitem{unpub}
See Section 11 and corresponding references in H. Rosu, 
``Hawking-like and Unruh-like effects: Toward experiments?",
Gravitation \& Cosmology {\bf 7}, 1 (2001) [gr-qc/9406012].

\bibitem{Frank}
I.M. Frank, Izv. Akad. Nauk SSSR, Ser. Fiz. {\bf 6}, 3 (1942).

\bibitem{bob}
B.M. Bolotovski and V.P. Bykov, 
``On the theory of ADE",
Radiofizika {\bf 32}, 386 (1989).

\bibitem{Marvin}
M.E. Goldstein, {\em Aeroacoustics}, (McGraw-Hill, New York, 1976).

\bibitem{bd} 
A.O. Barut and J.P. Dowling, 
``Quantum electrodynamics based on self fields:
On the origin of thermal radiation detected by an accelerated observer",
Phys. Rev. A {\bf 41}, 2277 (1990). 

\bibitem{eo}
D.H. Austin, K.P. Cheung, J.A. Valdmanis, and D.A. Kleinman, 
``Cherenkov radiation from femtosecond optical pulses in electro-optic 
media",
Phys. Rev. Lett. {\bf 53}, 1555 (1984). 
\\
See also,
T.E. Stevens, J.K. Wahlstrand, J. Kuhl, R. Merlin, 
``Cherenkov radiation at speeds below the light threshold: 
Phonon-assisted phase matching", 
Science {\bf 291}, 627 (2001) and references therein.

\bibitem{kysv}
See, \eg, 
V.V. Kurin, A.V. Yulin, I.A. Shereshevskii, and N.K. Vdovicheva,
``Cherenkov radiation of vortices in a two-dimensional annular 
Josephson  junction", 
Phys. Rev. Lett. {\bf 80}, 3372 (1998).

\end{thebibliography}
\end{document}